\documentclass{ws-procs9x6}

\begin{document}

\title{mSUGRA Dark Matter and the b Quark Mass}

\author{M.~E. G\'omez\footnote{\uppercase{S}peaker.}}

\address{ Dept. de F\'{\i}sica Aplicada, Universidad de Huelva, 21071 Huelva.\\
E-mail: mario.gomez@dfa.uhu.es}

\author{T.~ Ibrahim}

\address{ Dept. of  Physics, Faculty of Science,
University of Alexandria, \\
Alexandria, Egypt\footnote{\uppercase{P}ermanent address of\uppercase{ T.~I.}}\\ 
Dept. of Physics, Northeastern University,
Boston, MA 02115-5000, USA.\\ 
E-mail: tarek@neu.edu}  
\author{P. ~ Nath}

\address{Dept. of Physics, Northeastern University,
Boston, MA 02115-5000, USA.\\ 
E-mail: nath@neu.edu}  

\author{S.~ Skadhauge} 

\address{Dept. de F\'\i sica, Instituto Superior T\'ecnico, 1049-001 Lisboa, Portugal.\\
E-mail: solveig@cfif.ist.utl.pt}
\maketitle

\abstracts{We extend the commonly used mSUGRA framework to allow complex soft
terms. We  show how these phases can induce large
changes of the SUSY threshold corrections to the b quark mass and
affect  the neutralino relic density predictions of the model.
We present some specific models with large SUSY phases
which can accommodate the fermion electric dipole moment constraints and
a neutralino relic density within the WMAP bounds.}

\section{Introduction}
The recent Wilkinson Microwave Anisotropy Probe 
(WMAP) data  allows a determination of cold  dark matter (CDM) to
lie in the range\cite{bennett} 
$\Omega_{CDM} h^2 =0.1126^{+0.008}_{-0.009}$. In this analysis we extend
the mSUGRA framework to include CP phases in the gaugino 
sector \cite{our}, which affects the loop corrections to the 
b quark mass and also affects the mixing between 
 the neutral Higgs bosons. These corrections then affect 
 relic density computations in important ways.

\begin{figure}[hb]
\vspace{-1truecm}
\centerline{\epsfxsize=2.7in\epsfbox{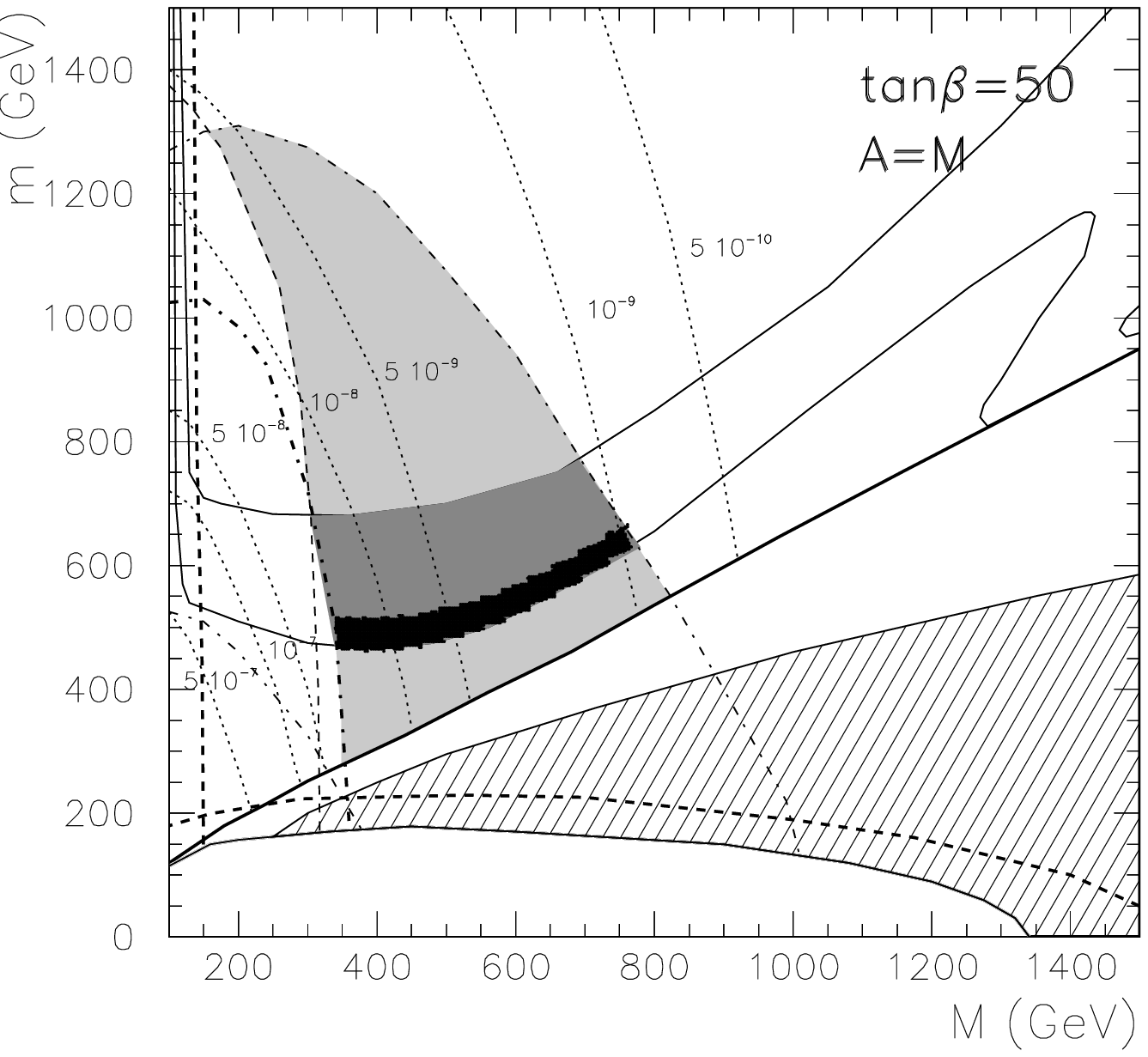},\hspace{-1truecm}\epsfxsize=2.7in\epsfbox{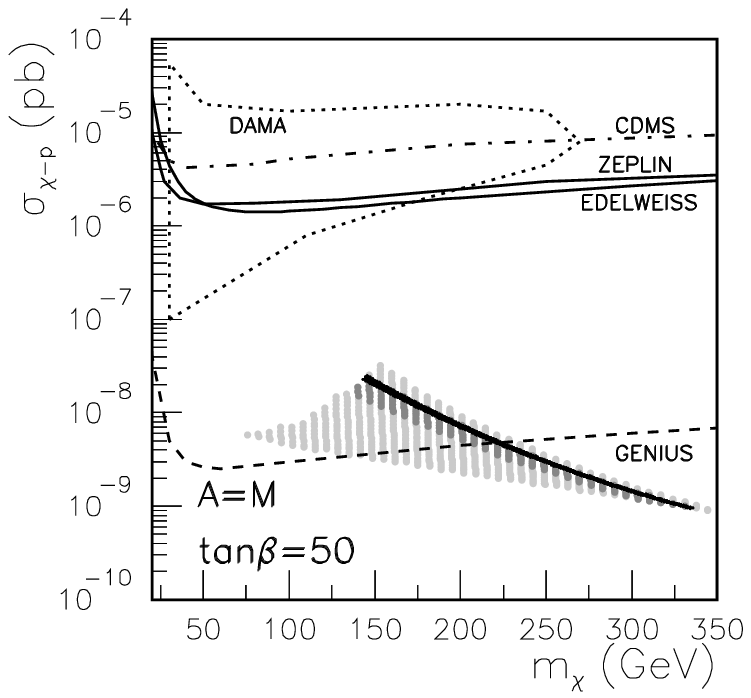} }
   \vspace{-.4truecm}
\caption{Neutralino relic density (left) and $\sigma_{\chi-p}$ 
(right) in mSUGRA at $\tan\beta=50$. Darker areas are favored by the phenomenological 
constraints described in Ref.~[7]. 
\vspace*{-13pt}
\label{msugra}}
\end{figure}

\section{mSUGRA Dark Matter}
In most of the mSUGRA parameter space \cite{msugra} the LSP is almost 
purely  a Bino $\tilde{B}$ with a large relic density. However, we can classify three 
regions where $\Omega_{\chi}$ can reach the WMAP bounds:
(i) Coannihilation region \cite{coan}: Relic abundance decreases due to 
coannihilations $\chi-\tilde{\tau}$ when $m_{\tilde{\tau}} \simeq m_{\chi}$. 
(ii) Hyperbolic Branch/Focus-point (HB/FP) region\cite{ccn}: The $\mu$--term is small, such that $\chi_{0}$ 
may have large Higgsino fraction which enables a faster annihilation.
(iii) Resonances on Higgs mediated channels\cite{Lahanas}: Relic abundance constraints are 
satisfied by annihilation though resonant  s-channel Higgs exchange.
In Fig.~\ref{msugra}, we present some representative  
mSUGRA predictions for  $\Omega_\chi h^2$ 
at large  $\tan\beta$\cite{cerdeno}. In the next section we analyze the impact 
of enlarging 
this picture  including CP phases using the following point in the point in the
mSUGRA parameter space
\begin{equation}
\tan\beta=50,\;m_0=m_{1/2}=|A_0|=600 \,{\rm GeV}. 
\end{equation}

\begin{figure}[t]
\vspace{-3.7truecm}
\centerline{\epsfxsize=2.5in\epsfbox{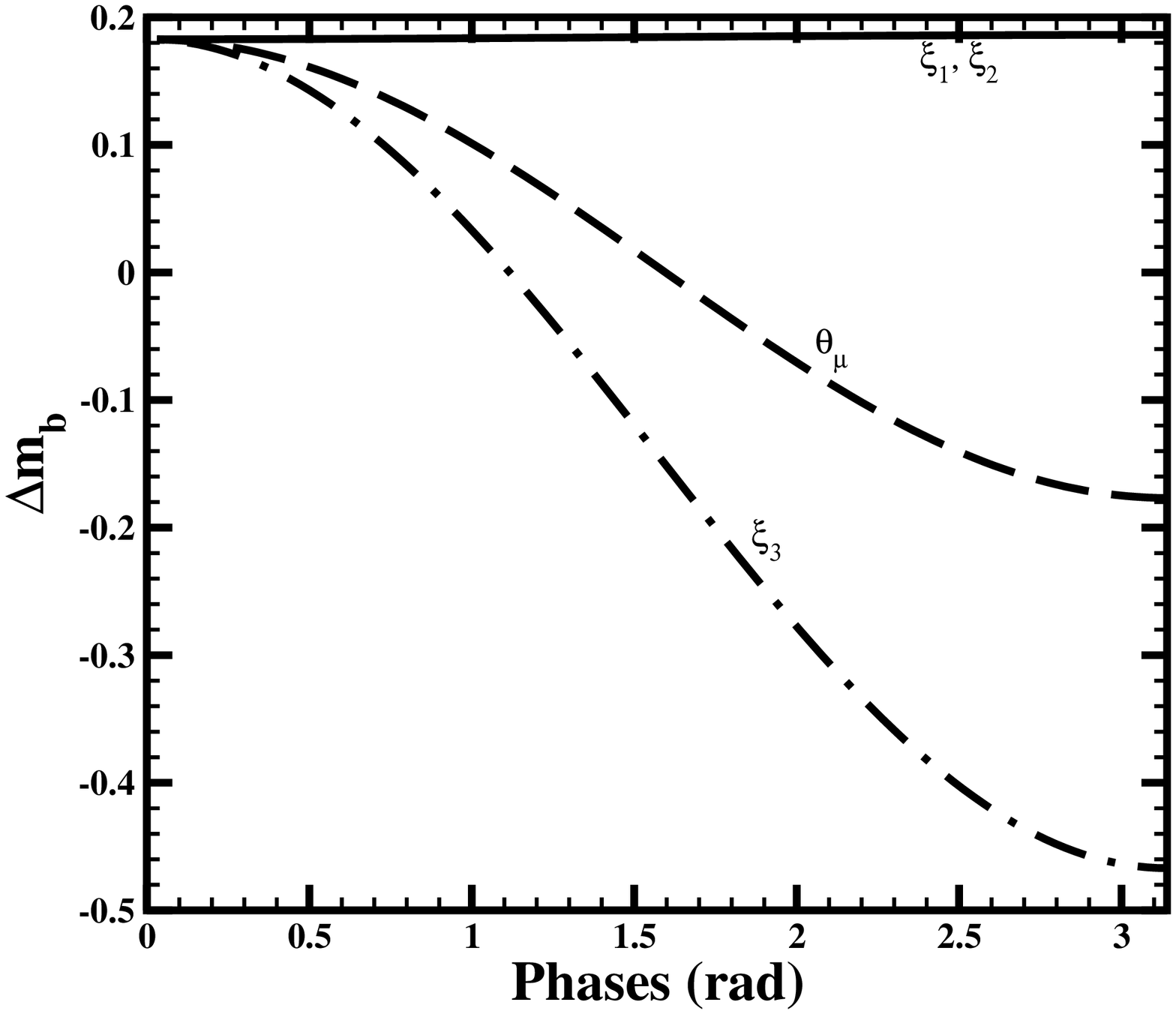},\hspace{-.5truecm}
\epsfxsize=2.5in\epsfbox{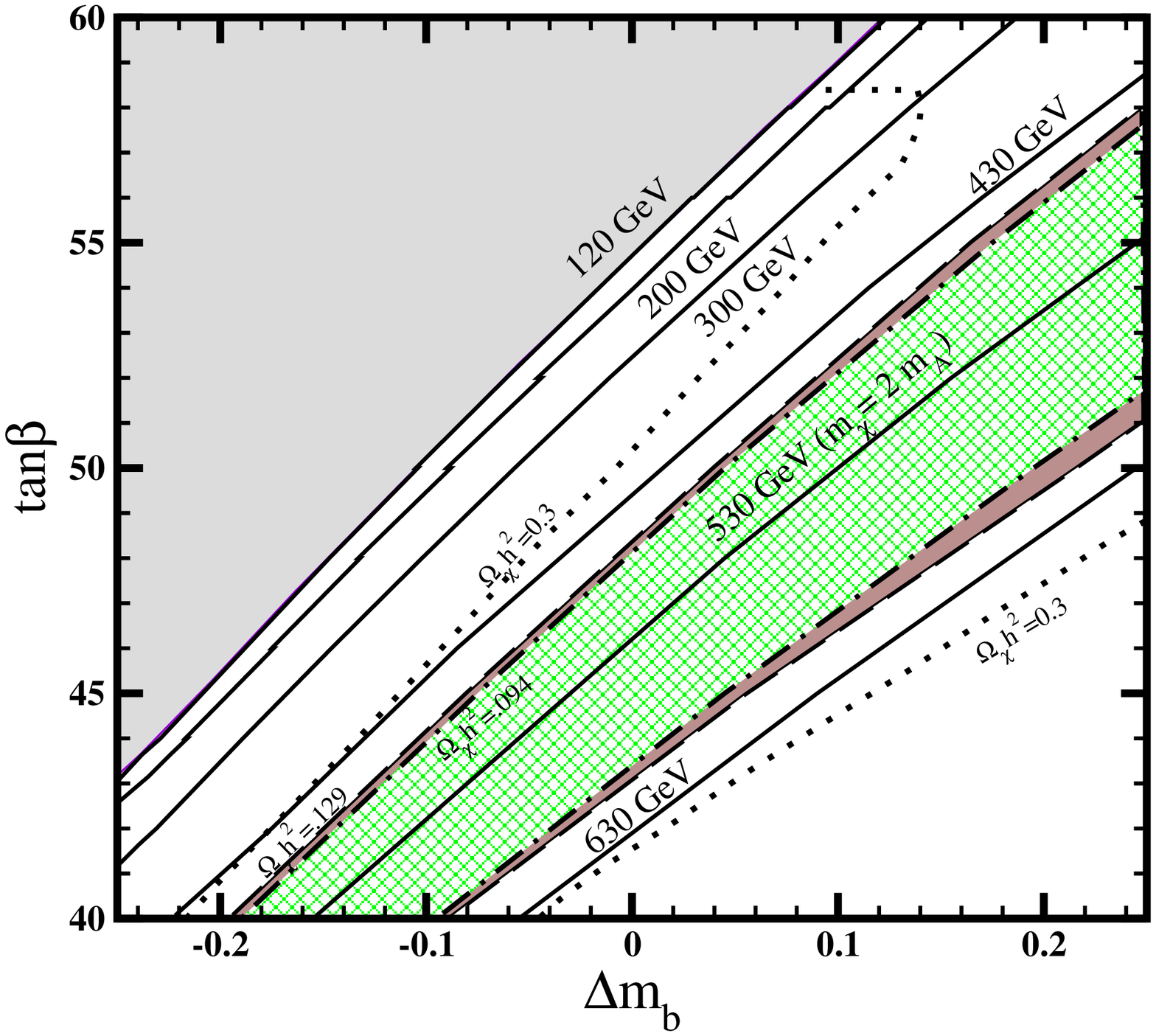} }
   \vspace{-.7truecm}
\caption{$\Delta m_b$ as a function of the indicated phases (left) and regions 
allowed by the relic density constraints for the parameters on Eq.~(1) of the text.
\vspace{-13pt} 
\label{delta}}
\end{figure}

\section{Phase Generalized mSUGRA}
Within mSUGRA there are only two physical phases,  $\theta_\mu, \theta_A$
 which are phases of $\mu$ and $A_0$. These phases must be small 
($\leq 10^{-2}$) to satisfy the electric dipole moments (EDM) constraints 
\begin{equation}
|d_e|< 4.23 \times 10^{-27}{\rm ecm},\;\; |d_n|< 6.5 \times 10^{-26}{\rm ecm}, \;\; C_{\rm Hg} < 3.0 \times 10^{-26}{\rm cm}.
\label{eq:EDM}
\end{equation}
Large phases can be accommodated in several scenarios such as models with 
{\it super heavy  sfermions} for the two first generations \cite{na} or models 
with a 
non--trivial {\it soft term} flavor structure\cite{Abel}. Here, we assume a 
cancellation 
mechanism \cite{cancel} which becomes possible if we assume an extended
SUGRA parameter space characterized by the parameters 
\begin{equation}
 m_0, m_{1/2}, \tan\beta, |A_0|, \theta_\mu, \alpha_A, 
\xi_1, \xi_2, \xi_3, 
\end{equation}
\noindent 
where, $\xi_i$ is the phase of the gaugino mass $M_i$. The value of 
$|\mu|$ is determined by imposing electroweak symmetry breaking.

\begin{figure}[t]
\vspace{-3.7truecm}
\centerline{\epsfxsize=2.5in\epsfbox{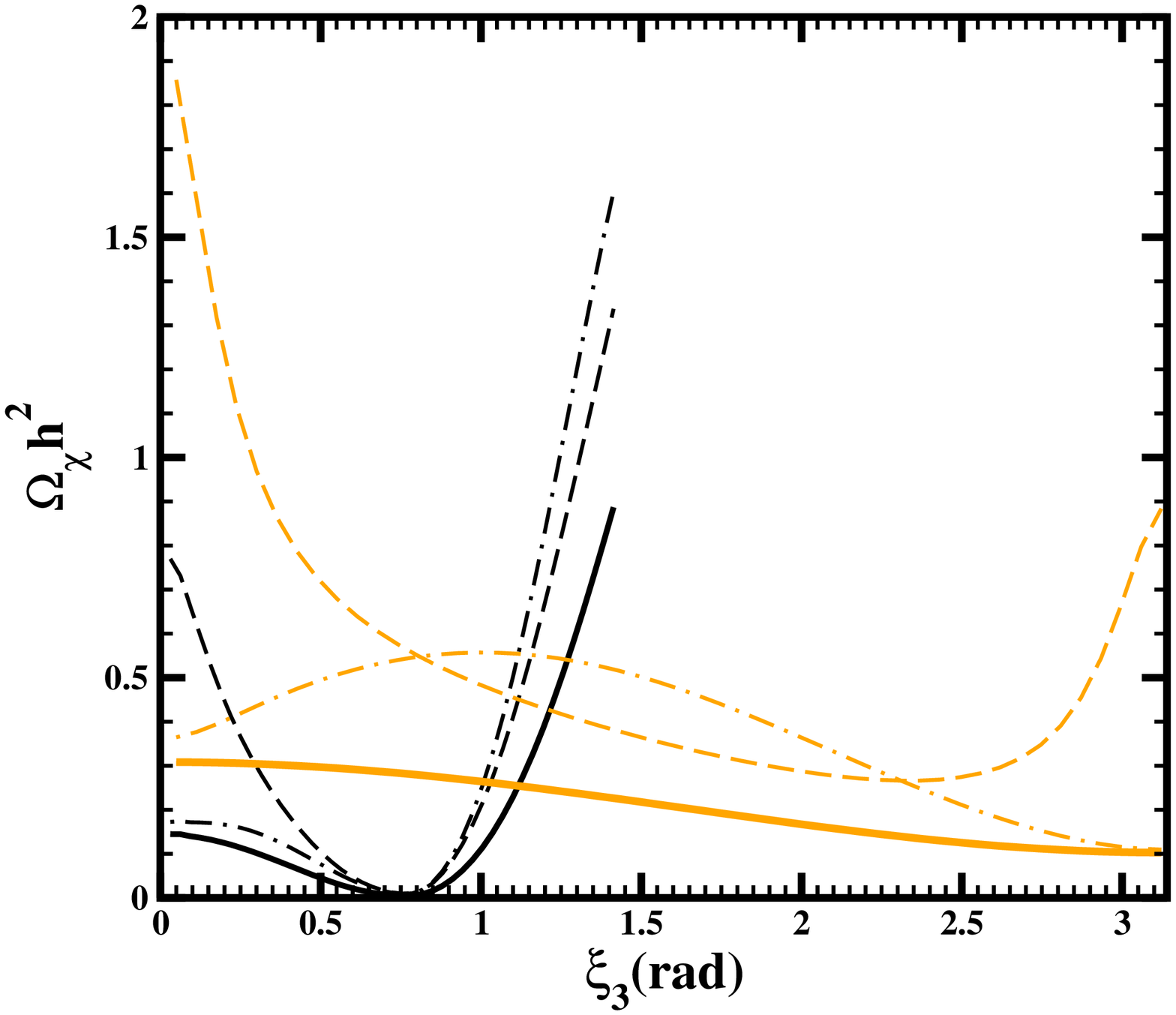}, \hspace{-.5truecm} 
\epsfxsize=2.5in\epsfbox{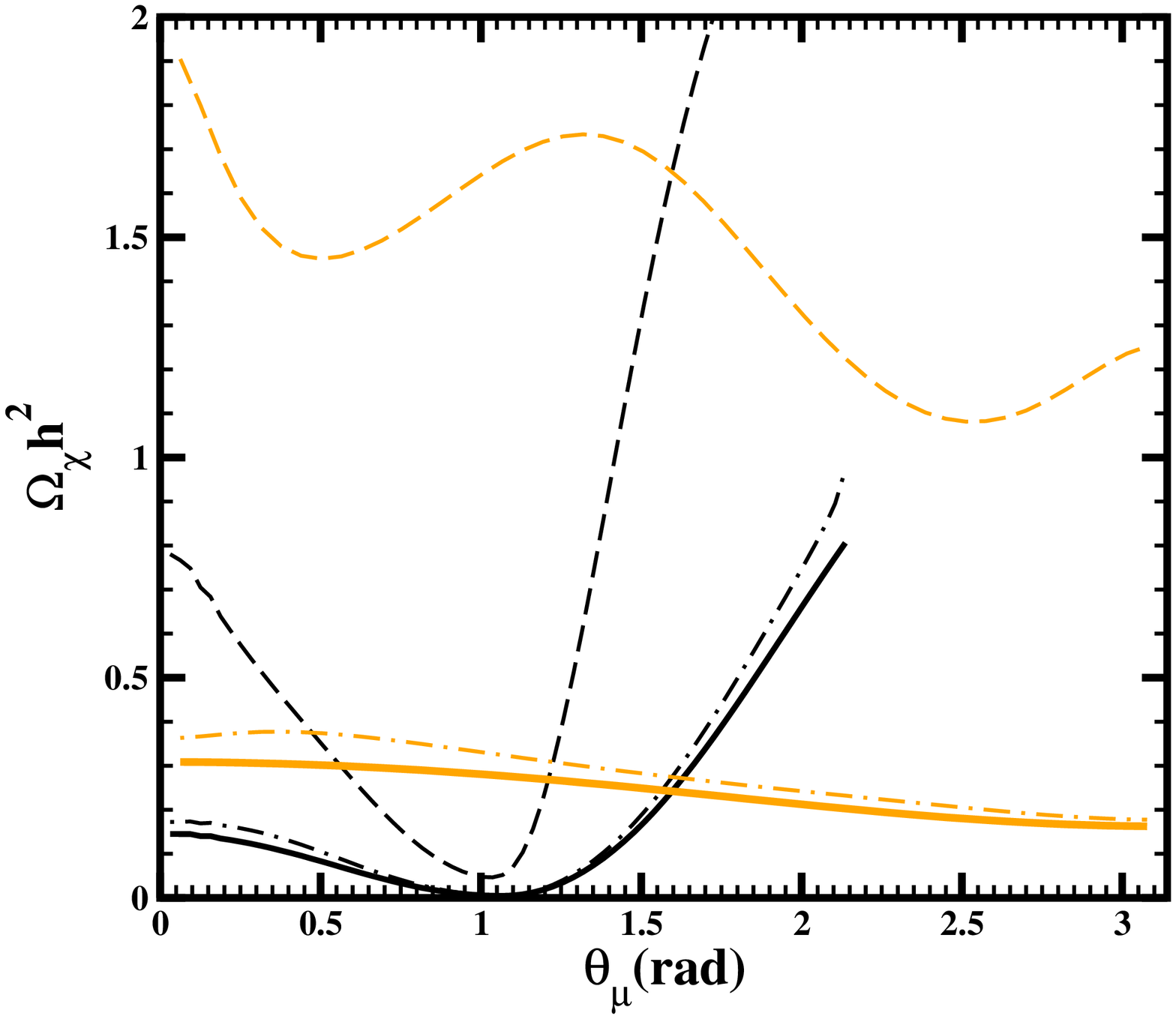} }
   \vspace{-.5truecm}
\caption{ $\Omega_{\chi}h^2$ as a function of $\xi_3$ and $\theta_{\mu}$ 
for the parameters on Eq.~(1), using the theoretically predicted value of 
$\Delta m_b$ (black lines), $\Delta m_b=0$ (light lines). Solid lines 
include all contributions, dashed lines (dot-dashed lines) 
only s-channel $H_1$ ($H_3$) mediated annihilation 
to $b\bar{b}$. 
\vspace*{-10pt}
\label{omega}}
\end{figure}

\subsection{Loop Correction to the b Quark Mass}
 At the loop level the effective b quark coupling with
the Higgs is given by\cite{carena2002} 
\begin{equation}
-L_{bbH^0}= (h_b+\delta h_b) \bar b_R b_L H_1^0 + 
\Delta h_b \bar b_R b_L H_2^{0*} + H.c.
\end{equation}
 The correction to the b quark mass is then given 
directly in terms of $\Delta h_b$ and $\delta h_b$ so that  
\begin{equation}
\Delta m_b= [Re(\frac{\Delta h_b}{h_b}) \tan\beta 
+Re(\frac{\delta h_b}{h_b}) ].
\end{equation}
A full analysis of $\Delta m_b$ is  used\cite{tarek}. $\Delta m_b$ depends strongly on 
$\xi_3$ and $\theta_\mu$ and weakly on $\alpha_A,\, \xi_1,\, \xi_2 $ as we  
see from the left panel of Fig.~\ref{delta}.  The consequences for 
$\Omega_\chi h^2$ arising from the changes of $\Delta m_b$ in this range can 
be understood from the qualitative analysis on the right 
panel of Fig.~\ref{delta}, where $\Delta m_b$ is used as a free parameter. We 
 observe that for a fix value of $\tan\beta$ the pseudo scalar Higgs mass 
can reach values in the range $m_A\sim m_\chi/2$, allowing predictions for 
$\Omega_\chi h^2$ \footnote{We use {\it micr0MEGAs} \cite{micro} for the computations 
of $\Omega_\chi h^2$ without phases. } on the WMAP bounds. 

\subsection{The Higgs sector CP--even CP--odd Mixing}
CP violating phases induce mixing at one loop of the CP--even, 
$H,h$, and CP--odd, $A$, neutral tree level  Higgs bosons:  
$(H,h,A) \rightarrow (H_1,H_2,H_3)$,  where  $H_i$ (i=1,2,3) are the mass eigen states.
The relevant couplings\footnote{
We use {\it CPsuperH} \cite{cpsuperh} in our  computations. }  for the 
s-channel neutralino annihilation become
\begin{equation}
{\it  L}= 
(S_k'+i S_k''\gamma_5)H_k\chi\chi
+(C^S_k +i C^P_k \gamma_5) H_k b \bar b. 
\end{equation}
The resonant annihilation cross-section behaves as
\begin{equation}
\sigma_{\rm ann} \sim 
\frac{(C^S_k C^S_l +C^P_k C^P_l) \left[S'_k S'_l 
(1-4 m_\chi^2/s)+S''_k S''_l\right]}{(-m_{H_k}^2+s-i m_{H_k}
\Gamma_{H_k})(-m_{H_l}^2+s+i m_{H_l}\Gamma_{H_l})}s^2 
\end{equation}
We  observe that imaginary couplings ($S''$) will dominate, since the 
real ones are suppressed by the factor $1-4 m_\chi^2/s$. Fig.~\ref{omega} shows the
 changes of $\Omega_\chi h^2$ with $\xi_3$ and $\theta_\mu$. The long 
light lines 
are obtained by setting $\Delta m_b=0$, which implies that the Higgs masses 
remain almost constant along these lines and hence we  see the 
effects of the CP phases on the vertices without the variation 
due to $\Delta m_b$. Also, $m_{H_1} \sim m_{H_3}$ and 
$\Gamma_{H_1} \sim \Gamma_{H_3}$, therefore large mixing are possible as 
we  see in the partial contributions of $H_1$ (dash) and 
$H_3$ (dot--dash) mediated s--channels.

\begin{figure}[t]
\vspace{-3.7truecm}
\centerline{\epsfxsize=2.5in\epsfbox{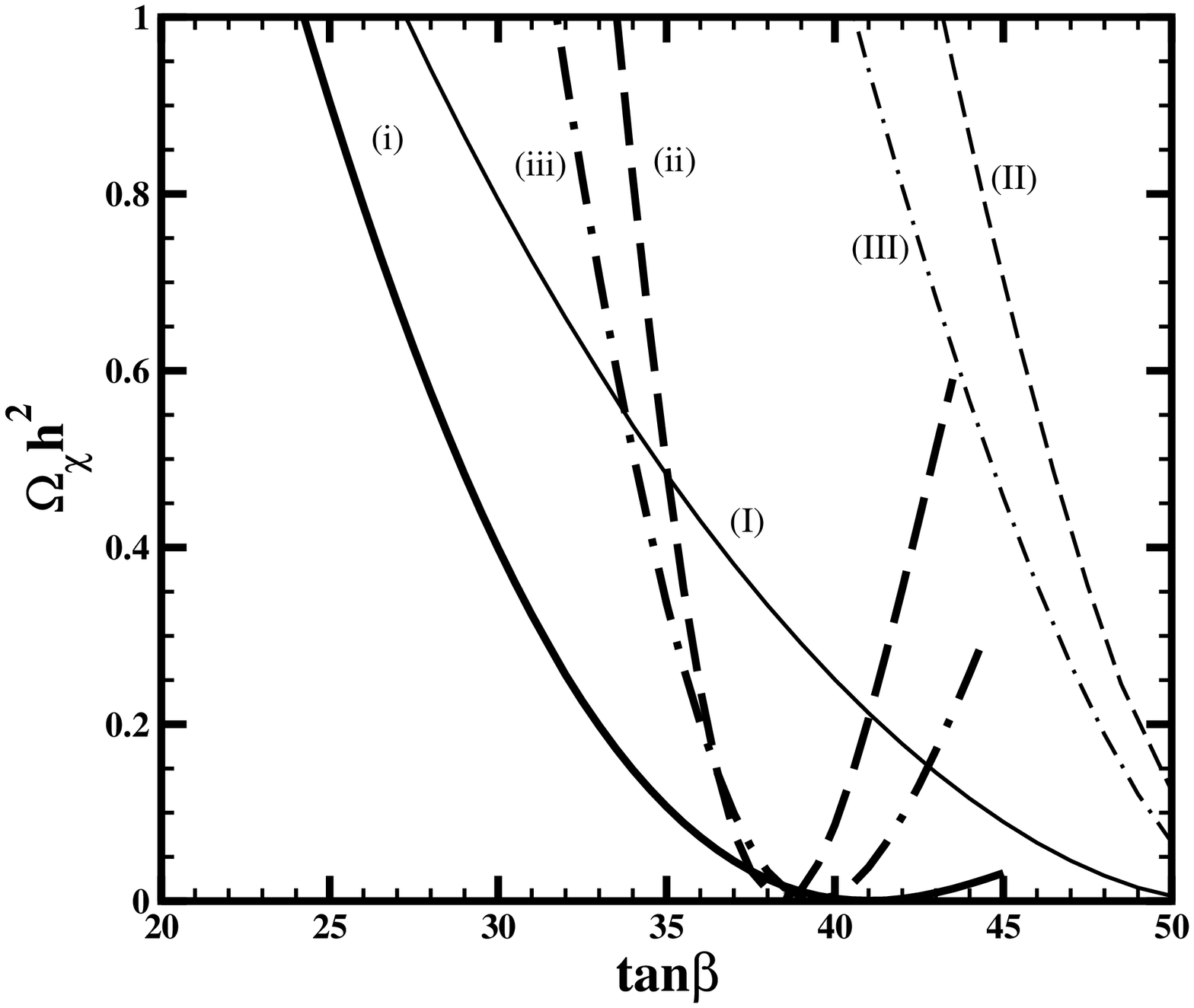}, \hspace{-.5truecm} 
\epsfxsize=2.5in\epsfbox{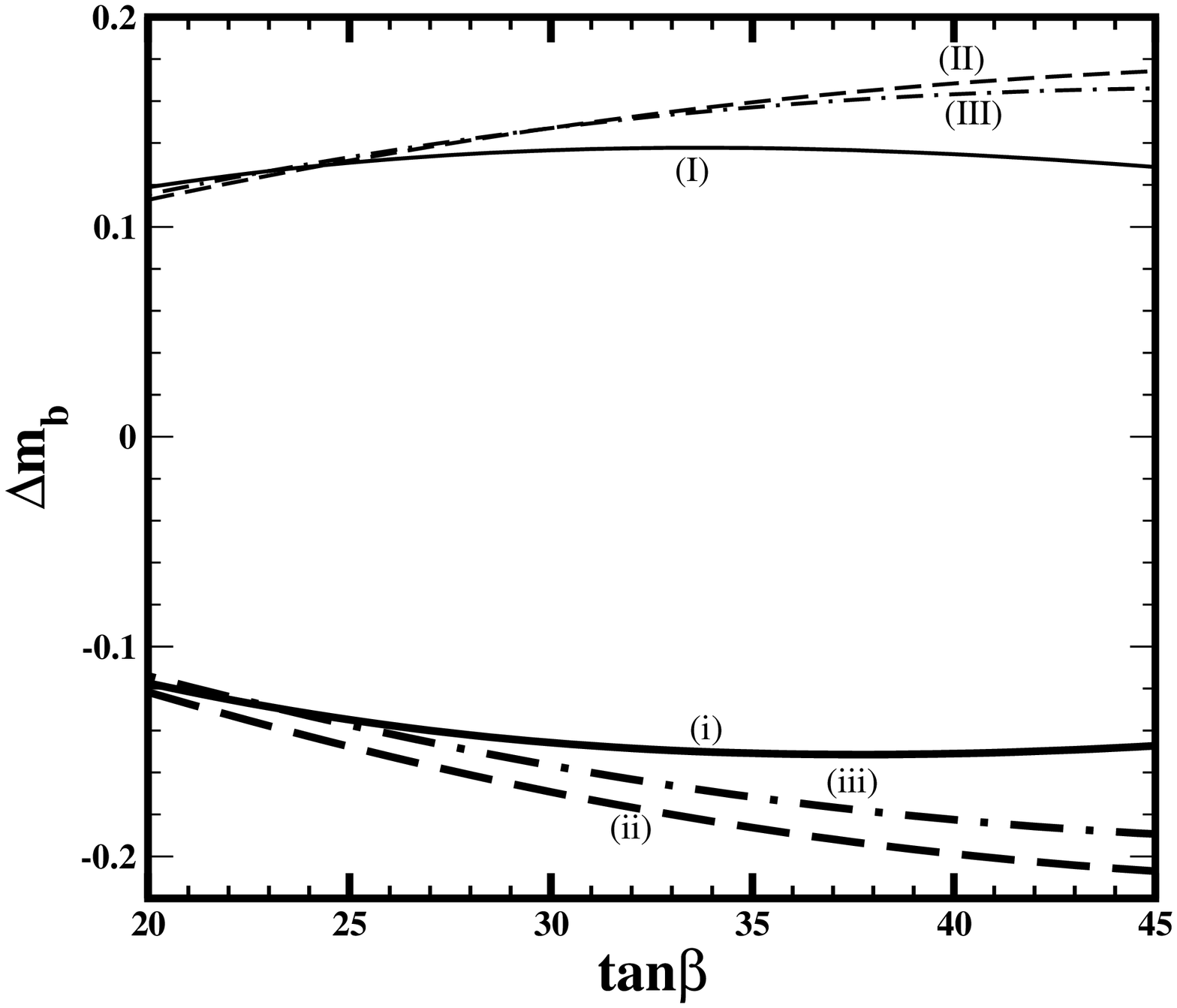} } 
 \vspace{-.5truecm}   
\caption{The neutralino relic density as a function of $\tan\beta$ for 
the three cases (i), (ii), (iii) of the text (left). Lines (I), 
(II) and (III) correspond to similar set of SUSY parameters for the case 
of vanishing phases. On the right we present the corresponding values 
of $\Delta m_b$. \label{EDM}}
\vspace*{-12pt}
\end{figure}
\section{CP--Phases, EDM's and Neutralino Relic Density}
In Fig.~\ref{EDM} the neutralino relic density is 
displayed as a function of $\tan\beta$ for three cases given by: 
(i) $m_0=m_{1/2}$=$|A_0|=300$ GeV, 
$\alpha_{A_0}=1.0$, 
$\xi_1=0.5$, $\xi_2=0.66$, $\xi_3=0.62$, $\theta_\mu=2.5$;
(ii) $m_0=m_{1/2}=|A_0|=555$ GeV, $\alpha_{A_0}=2.0$, $\xi_1=0.6$, 
$\xi_2=0.65$, $\xi_3=0.65$, $\theta_\mu=2.5$;
\indent (iii) $m_0=m_{1/2}=|A_0|=480$ GeV, $\alpha_{A_0}=0.8$, $\xi_1=0.4$, 
$\xi_2=0.66$, $\xi_3=0.63$, $\theta_\mu=2.5$. 
In all cases the EDM constraints~(\ref{eq:EDM}) are satisfied for 
$\tan\beta=40$ and their values are exhibited in table~\ref{table1}. 
We also observe that the WMAP bounds are also satisfied in the range of 
$\tan\beta$ exhibited in Fig.4.

\begin{table}[h]

\tbl{The EDMs for $\tan\beta=40$ for cases of text.}
{\footnotesize
\begin{tabular}{|l|l|l|l|}
\hline
Case & $|d_e| e.cm$ &  $|d_n| e.cm$ &  $C_{Hg} cm$  \\ \hline
(i) & $2.74 \times 10^{-27}$ & $1.79 \times 10^{-26}$ & $8.72 \times 10^{-27}$ 
\\ \hline 
(ii) & $1.29 \times 10^{-27}$ & $1.82 \times 10^{-27}$ & $6.02 \times 10^{-28}$
\\ \hline 
(iii) & $9.72 \times 10^{-28}$
 & $4.19 \times 10^{-26}$ & $1.41 \times 10^{-27}$  
\\ \hline 
\end{tabular}\label{table1}}
\vspace*{-13pt}
\end{table}

\section{Conclusions}
The SUSY threshold correction to $m_b$, $\Delta m_b$, can induce large 
changes on the two heavier neutral Higgs bosons. For a given $m_\chi$ and certain values of $\Delta m_b$ the resonance 
condition $m_{H_i}\sim 2 m_\chi$ can be satisfied. This implies a 
neutralino relic density inside the WMAP bounds. $\Delta m_b$ is strongly 
dependent on the SUSY phases $\xi_3$ and 
$\theta_\mu$. Hence, these phases can drive $\Omega h^2$ to the WMAP 
region leaving the possibility of choosing the other phases such that 
a cancellation mechanism keeps the fermion EDM predictions below the current 
experimental bounds.
\vspace{-.3truecm}

\section*{Acknowledgments}
 MEG acknowledges support from the 'Consejer\'{\i}a de Educaci\'on de 
la Junta de Andaluc\'{\i}a' 
and the Spanish DGICYT under contract BFM2003-01266.
The research of TI and PN was supported in part by NSF grant PHY-0139967. 
SS acknowledges support from the European RTN network HPRN-CT-2000-00148.

\end{document}